# Thermal Casimir Force Imaging of Nonequilibrium Hot Electrons

Weikang Lu[1,2], Ziyi Xu[1,2], Hewan Zhang[1], Svend Age Biehs[3], Achim Kittel[3], Ludi Qin[1], Xue Gong[1], Huanyi Xue[1], Yanru Song[4], Zhengyang Zhong[1], Shiyou Chen[5], Kun Ding[1], Wei Lu[2,6,*], Zhenghua An[1,7,8,9,*]

1 State Key Laboratory of Surface Physics, Institute for Nanoelectronic Devices and Quantum Computing, and Key Laboratory of Micro and Nano Photonics Structures (Ministry of Education), Department of Physics, Fudan University, Shanghai, China.
2 School of Physical Science and Technology, ShanghaiTech University, Shanghai 201210, China.
3 Institut für Physik, Carl von Ossietzky Universität, D-26111 Oldenburg, Germany.
4 ShanghaiTech Material Device Lab, ShanghaiTech University, Shanghai 201210, China.
5 State Key Laboratory of ASIC and System, School of Microelectronics, Zhangjiang Fudan International Innovation Center, Fudan University, Shanghai, China.
6 State Key Laboratory of Infrared Physics, Shanghai Institute of Technical Physics, Chinese Academy of Sciences, Shanghai 200083, China.
7 Shanghai Branch, Hefei National Laboratory, Shanghai 201315, China.
8 Shanghai Key Laboratory of Metasurfaces for Light Manipulation, Shanghai 200438, China.
9 Zhangjiang Fudan International Innovation Center, Fudan University, Shanghai 201210, China.
Email: luwei@shanghaitech.edu.cn (W.L.); anzhenghua@fudan.edu.cn (Z.A.)

**Abstract**

The thermal Casimir effect, arising from fluctuating electromagnetic fields of thermally agitated charges, induces thermosensitive forces and presents a novel approach to detecting nanoscale hot electrons—elusive yet ubiquitous in modern miniaturized transistors. However, detecting thermal Casimir forces at the nanoscale remains extremely challenging due to the background like electrostatic force and quantum Casimir force. In this study, we present the first non-contact force measurement of hot electrons based on thermal Casimir effect. Using an atomic force microscope (AFM) with a dual-resonant tip, we achieve thermosensitive force detection of nonequilibrium hot electrons while effectively suppressing background thermo-insensitive forces, thereby distinguishing them from cold electrons. In silicon nanoconstriction devices, the measured thermal Casimir pressure achieves approximately 3 bar at a separation of 5 nm at an electron temperature of $\sim 10^3$ K. Our work introduces a novel methodology for hot electron nanothermometry and provides critical insights into the thermo-mechanical properties of post-Moore nanoelectronics.

Force interactions underpin our universe, driving scientific exploration and transformative technologies. In the 1940s, Hendrik Casimir predicted that quantum fluctuations between two conductive plates could generate attractive forces, offering a macroscopic glimpse into quantum mechanics[1]. These forces become significant in a variety of micromechanical or microelectromechanical systems at submicrometer scales[2], such as adhesion and friction between movable objects[3,4]. Substantial progress has been made in the precise measurement and tuning of Casimir forces to achieve repulsion[5], as well as in demonstrating emergent applications such as

Casimir levitation[6], quantum trapping[7], and contact-free nanomachinery[8]. Beyond these forces, thermo-force interactions—ranging from steam engines to modern solid-state phase transitions—have garnered significant attention. Notably, thermal fluctuations introduce a temperature-dependent component to Casimir forces, known as the thermal Casimir force[9–13]. However, research on thermal Casimir forces has mainly focused on regimes comparable to or above the thermal wavelength ($d \sim \lambda_T = \frac{\hbar c}{k_B T}$; where $\hbar$ is the reduced Planck constant, $c$ is the speed of light and $k_B$ is the Boltzmann constant) due to the separation-dependent ratio between thermal and quantum forces ($\sim d/\lambda_T$)[9,10]. This focus has left the measurement of thermal Casimir forces in the deep sub-thermal-wavelength nanoscale regime unexplored, as well as the potential for temperature sensing such as hot electron nanothermometry.

Hot electrons reach much higher effective temperatures than the hosting lattice in modern microelectronics, significantly affecting device performance and raising serious heat management concerns for future developments[14]. Optimizing nanoscale thermal management by understanding and controlling hot-carrier kinetics is essential for advancing post–Moore-era nanoelectronics[14,15]. In particular, understanding the dynamics of hot electrons and tailoring their behaviors in both spatial and momentum spaces is critical to achieving ideal device functionalities, such as sharp switching on/off operations in transistors that surpass the Boltzmann limit for sub-threshold swing[16] and achieving near-unity solar cell efficiency that exceeds the Shockley–Queisser limit[17]. Additionally, the excess energy of hot electrons, along with their diverse interactions, opens up a range of promising applications, such as hot luminescent light sources[18,19], broadband photodetectors[20], thermoelectric devices[21], and plasmon-enhanced photochemistry[22]. However, a convenient and noninvasive technique for nanoscale mapping of hot electrons remains elusive. Scanning thermal microscopy (SThM)[23], while effective for contact-based thermal imaging[24,25], is unsuitable for detecting embedded hot electrons due to their extremely low heat capacity[26], which is orders of magnitude lower than that of the lattice. Alternative methods, such as pump-probe spectroscopy[27,28] and ultrafast electron microscopy[29], provide transient insights but are often invasive and incapable of capturing steady-state conditions in operational devices.

Here, we apply a bimodal operation of an atomic force microscope (AFM) to detect forces generated by nonequilibrium hot electrons. By using sideband demodulation, we discriminate the thermal Casimir forces from parasitic background forces, thereby unambiguously distinguishing hot electron patterns from their cold electron counterparts. The experimental results reveal a clear distribution of hot electrons on a silicon nanoconstriction device, showing linear correlation with the elevated electron temperature reaching as high as $T_e \sim 10^3\ K$, directly measured via a scanning noise microscope (SNoiM) at terahertz frequency[30]. Our work provides valuable insights into hot electron effects and their thermo-mechanical influence in modern downscaled semiconductor transistors.

**Experiment setup for force detection**

The experimental setup is based on a commercial AFM system and utilizes the first two resonant frequencies of the cantilever ($f_1$, $f_2$) for bimodal operation[31–33] (see Methods and Supplementary Note 1). Figure. 1a presents a schematic of the setup. The probe is mechanically driven at its second resonance $f_2$ for feedback control and topographic imaging (Fig. 1e). Simultaneously, the silicon nanodevices are electrically excited using a square wave alternating between 0 V (OFF) and $V_{ac}$ (ON) at a modulation frequency of $f_m$. To ensure synchronization, the electrical excitation is triggered by the AFM controller, which shares the same clock with the cantilever drive signal. Force signals induced by electrical excitation are resonantly detected at $f_1$, enhancing the force sensitivity. This bimodal approach benefits from the higher stiffness of the second resonance ($f_2$), enabling stable operation at smaller tapping amplitudes ($A_2$) and facilitating nanoscale separations. To precisely extract the force difference between the ON (hot electron) and OFF (cold electron) states, the modulation frequency is set to the difference between the two resonances ($f_m = \Delta f \equiv f_2 - f_1$), a method we term sideband modulation (SM) mode (Fig. 1c). This approach allows the measured signal to reflect the force difference between conditions where electrons are hot ($T_e \gg T_{lattice} \approx T_{tip} \approx 300\ K$) and cold ($T_e, T_{lattice}, T_{tip} \simeq 300\ K$), as will be seen later. We also note that electrical excitation during the ON state may introduce a non-negligible electrostatic potential change between the conducting channel and the electrically grounded tip[34]. To account for this, control experiments were conducted with the modulation frequency matched to first resonance ($f_m = f_1$), referred to as direct modulation or DM mode (Fig.1d), to measure and compare parasitic electrostatic forces (See Supplementary Note 2). To avoid complex surface conditions, silicon samples were vacuum-baked to remove moisture after fabrication and immediately transferred to the measurement chamber (see Methods). All experimental conditions—aside from the biasing condition ($V_{ac}$), modulation frequency ($f_m = \Delta f$ or $f_1$), and tip height ($z$) —were held constant. This ensures that other thermo-insensitive forces, such as short-range van der Waals and vacuum Casimir forces, remained consistent across ON and OFF states and were effectively canceled through lock-in demodulation. To quantify the elevated electron temperature ($T_e$) during the ON state, the same bias amplitude was applied in the SNoiM measurements. The SNoiM acts as a near-field radiative thermometer, directly imaging the temperature of hot electrons[30,35–37] (see Methods and Supplementary Note 3).

**Distinguish electrostatic force and thermal Casimir force**

Figure 1d shows the force distribution measured in DM mode, where the force decreases monotonically from the source to the drain. In contrast, Fig. 1c (SM mode) reveals a distinct force peak at the constricted center of the device. This stark contrast between DM and SM modes indicates that different forces dominate in each demodulation mode. Figure 2 illustrates the force distribution under varying current directions (by exchanging source and drain electrodes). Reversing the current direction flips the force distribution in DM mode (Figs. 2a, b), as also shown by the antisymmetric line profiles (solid dots) in Fig. 2c. Finite-element simulations of the force profiles (solid lines in Fig. 2c), based on electrostatic potential distributions (Fig. S1c), align with the experimental data, indicating that the DM mode force is primarily electrostatic force $F_{es}$ with negligible contribution from thermal effects. In SM mode (Figs. 2d–f), the force peak remains fixed at the constriction, indicating its independence on current direction and suggesting a thermal origin of Joule heating mechanism. The baseline of the line profiles (Fig. 2f) shows a current-direction dependence similar to that in DM mode, attributable to a residual electrostatic force ($F_{es}$), which contributes $\sim 10 \pm 1\%$ of total force in Fig.2f (depending on the tip height, see Fig. S3). The observation of the thermal-related force in SM mode is attributed to the fact that, in this mode, the measured signal is force gradient[32,33,38], unlike in the DM mode the force itself is detected (Supplementary Note 2). As a result, SM mode is more sensitive to shorter-range force than DM mode. During measurements, high current density in the constricted region induces substantial thermal effect via Joule heating, enhancing local charge fluctuations and generating fluctuating electromagnetic evanescent fields on the sample surface[30]. These surface waves are essential to the short-distance physics of the Casimir force[39] and play a crucial role in the measured thermal force signal in our work (Figs. 2d-f). The fluctuating, near-field nature of thermal Casimir is in stark contrast to the long-range electrostatic force ($F_{es}$). This explains why thermal Casimir force is observed only in SM mode while $F_{es}$ prevails in DM mode but contribute less to SM mode.

We further examined the z-dependence of the measured forces in SM mode to distinguish contributions from different forces with characteristic ranges[40]. Specifically, we measured force-distance curves at three representative positions (A, B, C) along the channel, spaced approximately 1μm apart, as shown in the inset of Fig. 3a. As the tip approaches the sample surface, the measured force initially increases, then decreases, forming a peak at $z \approx 8.2$ nm for positions A (Fig. 3b) and C (Fig. 3d). This peak arises because both tapping amplitude and effective quality factor decrease once the oscillating tip enters the force range of the sample surface (Fig. 3a, also see Supplementary Note1). In contrast, for the middle position B, the force curve shows two peaks (Fig. 3c). One peak corresponds to the same feature observed at positions A and C, while its intensity decreases from A to B and then C, in agreement with the trends of $F_{es}$ observed in Figs. 2a-c. We infer that this peak (labeled as Peak $R$) originates from $F_{es}$. The other peak appearing at a shorter distance ($z \approx 4.9$ nm, Peak $L$) is primarily due to the current-direction-independent thermal force signal ($F_{th}$) observed in Figs. 2d-f. This result highlights the importance of selecting a sufficiently close tip-sample distance to detect the thermal-dominant force signal, which is different from conventional larger distance regime for thermal Casimir forces studies, typically on the order of or above the thermal wavelength ($\lambda_T \approx 7\ \mu m$ for 300 K). Noting that thermal Casimir force is expected to follow the scaling relation $\frac{\partial F_{th}}{\partial z} \propto \frac{1}{z^\eta}$ (or $F_{th} \propto \frac{1}{z^{\eta-1}}$), we extract the power index $\eta$ for $F_{th}$, with the zero-separation point carefully calibrated using a point-mass harmonic oscillator model[41,42] (see details in Supplementary Note 5). For comparison, we find that the electrostatic force gradient exhibits a power index $\eta = 2.06 \pm 0.05$, aligning well with theoretical expectation (green dash line in Fig. 3e) for a sphere-plate configuration[34,43], confirming the reliability of our method. Figure 3e also displays the theoretical values: $\eta = 7$ for van der Waals forces between molecules, and $\eta = 3$, $\eta = 4$ respectively for thermal and quantum Casimir forces under ideal conditions[9]. The experimentally extracted $\eta$ values for the purely thermal Casimir component (after subtracting the residual electrostatic background) are slightly above the expected thermal value ($\eta = 3$) but below the quantum value ($\eta = 4$).

**The hot electrons induced thermal force**

To distinguish whether the measured thermal force signals are ascribed to hot-electron- or lattice-subsystems in a non-equilibrium state[30,35–37], we highlight the following key points: First, the thermal force signal is observed exclusively within the constricted mesa region of the device (Figs. 2d, e) and not in the surrounding areas where the doped silicon layer has been etched. This suggests that electrons play a crucial role in the detectable force signal, as lattice heat would diffuse readily into the etched regions rather than remain localized[35,36]. Second, generating a substantial thermal force signal requires a significant temperature difference between the tip and the sample. However, the increase in lattice temperature ($T_L$) is inherently constrained by material properties, such as thermal capacity and melting point. In contrast, under external biases, the electron subsystem can achieve a steady state with exceptionally high $T_e$, far exceeding lattice temperature ($T_e \gg T_L$)[30,35–37]. Experimental measurements of the effective electron temperature ($T_e$) in the constricted region, conducted using a home-built SNoiM[30] (Supplementary Note 3) reveal a strong correlation between

$T_e$ and measured force signal. As shown in the inset of Fig. 4a, the hotspot at the channel center shows a typical peak $T_e$ of approximately 965 K under an excitation bias of $V_{ac} = 4\,V$, consistent with previous studies[35]. The variation of $T_e$ with different biases ($V_{ac}$) follows a quadratic dependence, as predicted by Joule's law (see Fig. S5 and also Supplementary Note 3 for details). In Fig. 4a, line profiles of the thermal force and $T_e$ show strong alignment, with nearly identical peak positions and widths. Furthermore, Fig. 4b plots the thermal force gradient against $T_e$, revealing a linear dependence with a slope of $\frac{1}{\Delta T_e}\frac{\partial F}{\partial z} \sim 2.15 \times 10^{-4}\,N/(m \cdot K)$. This linear dependence provides strong evidence that hot electrons dominate the observed thermal force. Third, the peak change in lattice temperature, experimentally measured by SThM, is $\Delta T_L \sim 3\,K$ (Fig. S9b), more than two orders smaller than the electron temperature change $\Delta T_e \sim 665\,K$ for $V_{ac} = 4\,V$, consistent with earlier reports[35,36]. This implies that the lattice-induced thermal Casimir force is negligible compared to the electron-induced counterpart. Additionally, the permittivity change due to $\Delta T_L \sim 3\,K$ is estimated to be $\sim 0.3‰$[44], resulting in a Casimir force change by an order of magnitude lower than the contribution from hot electrons. Besides, the lattice expansion corresponding to $\Delta T_L \sim 3\,K$ is only ~0.6 pm, far below the instrumental noise level of feedback z-control, ruling-out the possibility of its contribution to the force signal in Figs. 3b-d (see also Supplementary Note 6).

Existing theories, such as Lifshitz theory[11], are not directly applicable to our nonequilibrium conditions ($T_e \neq T_L$) or the nanoscale separation regime with comparable surface roughness ($\delta z = 4.4\,nm \sim z$, Fig. S11a) and tip radius ($R \cong 30\,nm$). Hence, electron contributions are evaluated within a wide uncertainty range, $1.3 \times 10^{-8} \sim 5.4 \times 10^{-5}\,N/(m \cdot K)$ corresponding to artificially adjusted separations ($z + \delta z$ or $z - \delta z$) in order to take account of large roughness. Experimentally, the measured forces are close to the upper limit of this range, consistent with reports that surface roughness in short-distance regimes can significantly increase the force[45] (Fig. S11b). The remaining factor of ~4 difference between the experimental value and the theoretical upper boundary lacks a definitive explanation. It might stem from inaccuracies in calibrating the cantilever's spring constants, systematic errors potentially underestimating $T_e$, surface roughness etc. These findings may stimulate theoretical interest in studying thermal Casimir forces in ultra-small distance regimes, going beyond traditional fluctuation-dissipation theorem and perturbative corrections for roughness[2]. Early observations of anomalously large heat flux in nanoscale gaps further underscore the need for exploration in this area[46]. Future efforts would require improved roughness (e.g., atomically flat sample surface) and more general theoretical frameworks, incorporating practical considerations such as large roughness, $T_e$-dependent Drude parameters[47] of hot electrons and softening effects[48] etc.

While precise measurements of absolute thermal Casimir forces are beyond the scope of this study, our technique unambiguously demonstrates the relative distribution of electrically induced thermal Casimir forces. It is interesting to note that the experimental setup parallels the configuration of modern transistors, where the gate electrode and current-carrying channel correspond to the AFM tip and the silicon nanoconstriction channel in present work. At a tip-sample separation of $z = 5\,nm$, the measured force is effectively exerted on a spot with a radius $r \cong \sqrt{2\pi z R} = 17\,nm$ (see Supplementary Note 4 for details). The effective thermal Casimir pressure reaches $\sim 3.08 \times 10^5\,Pa$ (~3 bar), corresponding to a Casimir field of $\sim 2.14 \times 10^3\,V/cm$. For smaller distance such as $z = 2\,nm$ (the typical thickness of the oxide layer in modern devices[49,50], the thermal Casimir pressure will increase to $\sim 8.34 \times 10^6 \, Pa$ ($\sim 83 \, bar$) and the Casimir field reaches $\sim 5.79 \times 10^4 \, V/cm$, a significant portion of source-drain biasing field. These findings highlight the significance of thermal Casimir forces in nanoscale devices of post-Moore era.

**Conclusion**

We have demonstrated hot electron mapping by detecting the thermal Casimir force using sideband demodulation. The measured force exhibits a clear linear dependence on $T_e$, as anticipated. Furthermore, we assess the impact of this force at scales relevant to the thickness of the oxide layer in modern transistors, highlighting its potential significance in optimizing device performance. In conclusion, our work presents an accessible approach for studying hot electrons in real devices, addressing critical challenges in industry development. Moreover, our experimental findings may inspire theorists to investigate the complex physics that arises in such narrow gaps.

**Method**

**Sample preparation**

The Si nanodevice is defined on phosphorus-doped ($\sim 10^{20}$ cm$^{-3}$) silicon films that are epitaxially grown (90 nm thick) on intrinsic Si substrate. A nano-constriction structure is patterned using electron beam lithography (EBL) with PMMA resist, where the narrowest region of the constriction is approximately 500 nm in width. The nanodevice is then created by reactive ion etching (RIE) to a depth of about 120 nm, ensuring that the conductive film outside the channel is completely etched away. After etching, the PMMA resist is removed using acetone. Before transferring the sample to the measurement chamber, samples were baked at approximately 120 °C in a high vacuum chamber (about $10^{-6}$ mTorr) for 48 hours to remove any contaminations such as moisture from the surface.

**Force measurement**

The force measurement experiment was conducted using a commercial AFM (Vista One, Molecular Vista Inc.). The excitation source at the modulation frequency $f_m$ was an arbitrary waveform generator (AFG3022C, Tektronix), which was triggered by the AFM controller in burst mode, with a duty cycle controlled at 10%. Different modulation frequencies were chosen to carry out the force measurements either in DM mode ($f_m = f_1$) or in SM mode ($f_m = f_2 - f_1$). All the force measurements were conducted in a dry nitrogen atmosphere. Two types of cantilevers were used. The first type, PPP-XYNCHR from Nanosensors, was coated with a 65 nm-thick gold layer. The second type, NSC18/Cr-Au from Micromasch, had a gold coating as well. The resonance frequencies for the first type were approximately $f_1 \approx 250$ kHz and $f_2 \approx 1.6$ MHz, while the second type had resonance frequencies of about $f_1 \approx 80$ kHz and $f_2 \approx 500$ kHz. The minimal detectable force is limited by thermal fluctuations of cantilever[51,52], which is given by $F_{min} = \sqrt{\frac{4k_B TBk}{\omega Q}}$, with Boltzmann's constant $k_B$, temperature $T$, spring constant $k$, detection bandwidth $B$, frequency $\omega$ and Q factor of the cantilever. Both types of tips have comparable $F_{min}$ ($\sim 55 \, fN/\sqrt{Hz}$ for the first type and $\sim 51 \, fN/\sqrt{Hz}$ for second) compared to the first type (about $55 \, fN/\sqrt{Hz}$). Both tips were found to give consistent imaging data implying robustness of our results.

**$T_e$ measurement with SNoiM**

A home-built Scanning Noise Microscope (SNoiM) is used for studying the electron temperature equipped with a cooled HgCdTe detector (center wavelength is $11\ \mu m$). Note that the phonon frequency of silicon is out of the spectral range of the detector, ensuring no contribution from hot phonons or lattices to SNoiM signal[30]. As a result, detected with SNoiM is the charge/current fluctuation from hot electrons that generates intense evanescent waves but cancels out in the region away from the surface[30]. In this work it is the hot-electron shot noise, the intensity of which is most simply characterized by the effective electron temperature $T_e$. During the experiments of Si nano-constriction device, an external pulsed bias is applied and the signal is demodulated with $f \approx 1.7\ kHz$. Absolute values of $T_e$ are derived from the signal intensity measured under different bias voltage. The principle and the construction of SNoiM are described in more details in refs.[30,36] and in Supplementary Note 3.

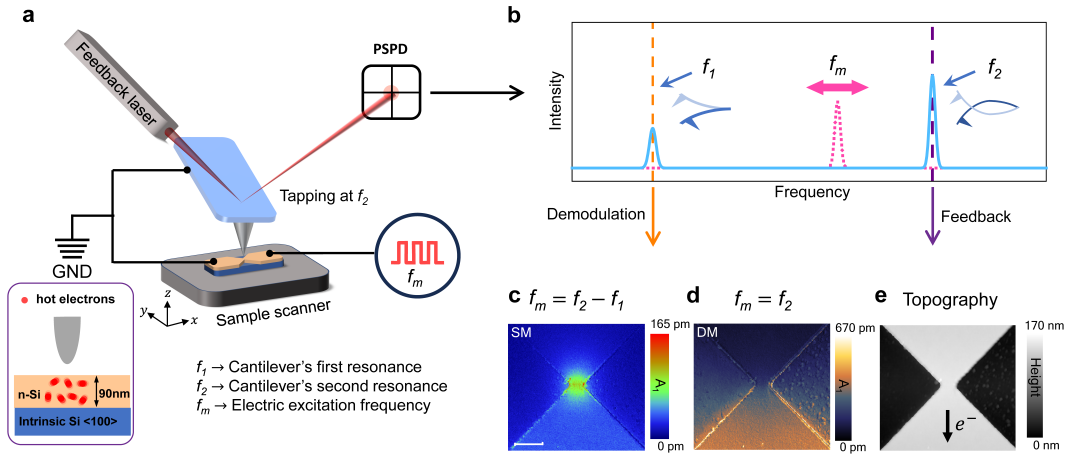

**Fig. 1 Bimodal AFM for force measurement. a** Sketch of the experimental setup. **b** Spectrum of cantilever motion. The blue solid line represents the cantilever's motion measured by the position-sensitive photodiode (PSPD), while the pink dashed line indicates the tunable electric excitation at the modulation frequency $f_m$. **c, d** Results of the sample excited by 4V-pluse demodulated at the first resonance frequency $f_1$ in SM (**c**) and DM (**d**) modes. **e** Topography of the sample, acquired with the feedback control at the second resonance frequency $f_2$. The scale bar for **c-e** is 1μm.

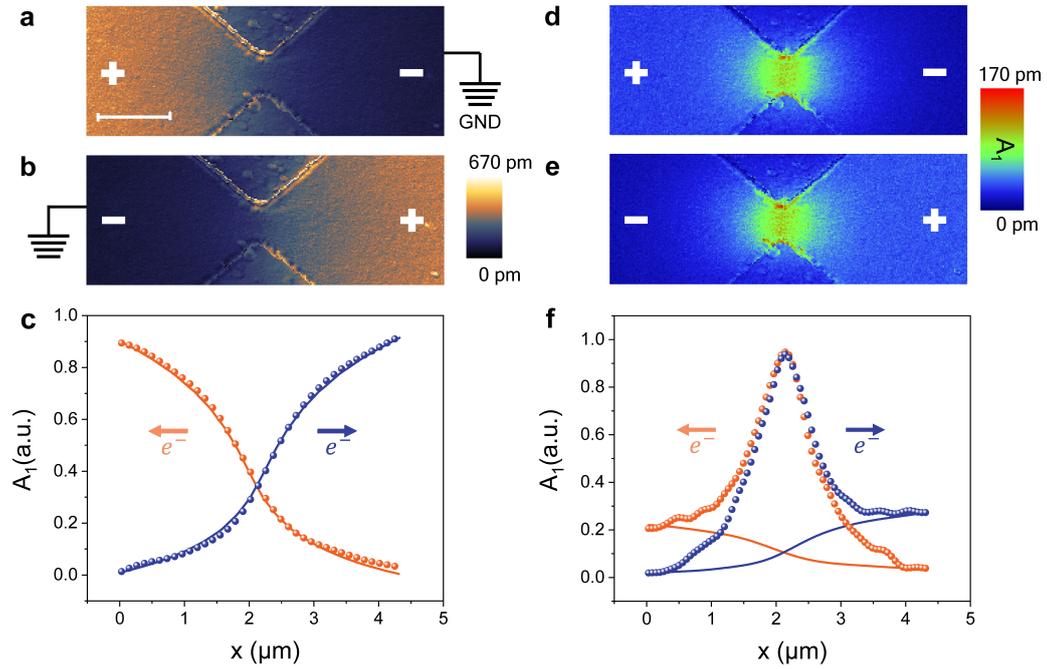

**Fig. 2 Force imaging in DM and SM modes with different current directions. a, b** Force mapping in DM for different current directions. **c** Line profiles in DM mode along the channel for different electron flow directions. The scatter points represent experimental data, while the solid lines are results from finite-element numerical simulations. **d, e** Force mapping in SM mode for different current directions. **f** Line profiles in SM mode along the channel for different electron flow directions. The scatter points represent experimental data, and the solid lines are electrostatic residuals obtained from DM results. The scale bar is 1μm.

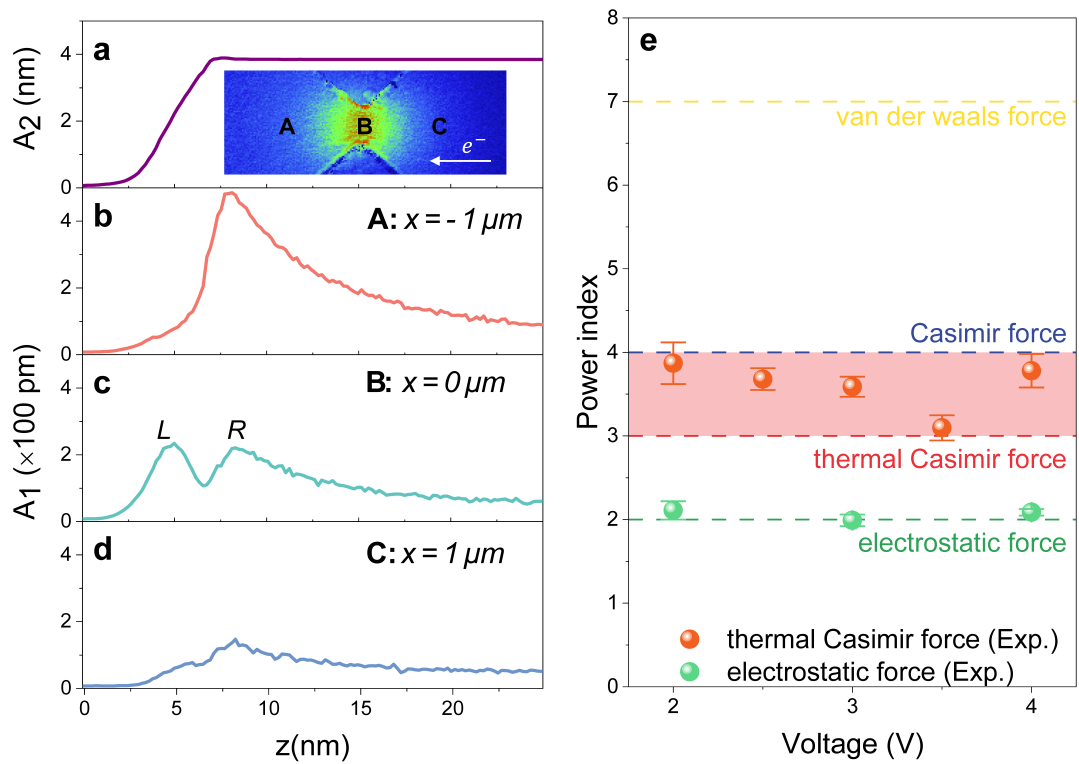

**Fig. 3 Separation dependence of forces measured in SM mode. a** Z-dependence curve of tapping amplitude $A_2$ during force curve measurement with $V_{AC} = 4V$ excitation. Inset: A,B,C label the positions with 1μm spacing along the channel and B, centered at the nanoconstriction. **b-d** Force curves obtained at position A, B and C. The tip-sample separation is calibrated by fitting tapping amplitude with theory model, as presented in Supplementary Note 5. Labels "*L*" and "*R*" represent the thermal force peak and electrostatic force peak, respectively. Each point is obtained from force curves under 16 times average and the single curve is done with about 40nm/s speed to reduce the influence of sample shifting with time. **e** Power index ($\eta$) of force gradient versus separation with different excitation voltages. Scatter points are experimental data. Dash lines with different colors show the theoretical references of different forces: electrostatic force (green), van der waals force between molecules (yellow), quantum Casimir force (blue) and thermal Casimir force (red).

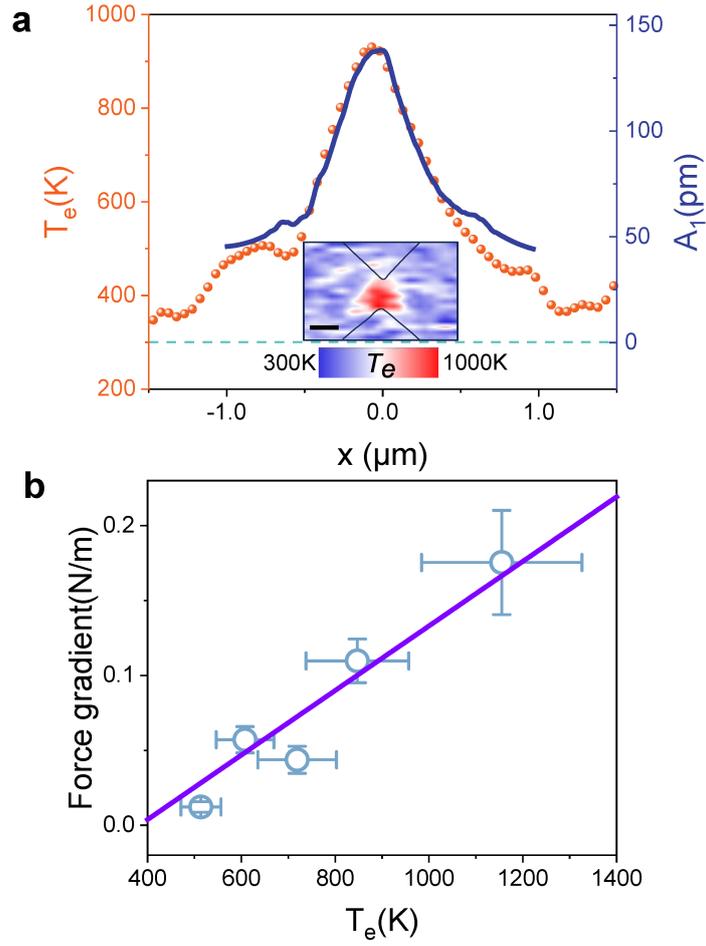

**Fig. 4 Correlation between $T_e$ and thermal force. a** Comparison of line profiles of $T_e$ and thermal force along the channel with 4V excitation. Orange dots are $T_e$ obtained by SNoiM and dark blue line represents thermal force. Inset is $T_e$-mapping and the scale bar is 500nm. **b** Linear fitting of thermal force gradient versus $T_e$. The transverse error bar represents 20% uncertainty of $T_e$ due to the signal noise ratio. The vertical error bar shows the standard deviation of thermal force of 1024 repeated readings with 10ms time constant. The fitting result is $\frac{\partial F(\Delta T_e)}{\partial z}\big|_{z=5nm} = (2.15 \times 10^{-4})\frac{N}{m \cdot K} \cdot \Delta T_e$.

**Data availability** The data that support the findings of this study are available from the corresponding authors upon reasonable request.

**Code availability** The code that used for calculating force and Hamaker function of this study are available from the corresponding authors upon reasonable request.

**Acknowledgements** Z.A acknowledges the financial support from National Key Research and Development Program of China (Grant No. 2024YFA1409800), Innovation Program for Quantum Science and Technology (Grant No.2024ZD0300103), the National Natural Science Foundation of China under Grant Nos. 11991060/12027805/12474042, Shanghai Science and Technology Committee under Grant No. 23DZ2260100, and the Sino-German Center for Research Promotion (No. M-0174). We thank Prof. Kun Ding for helpful discussions. W.K.L. acknowledges the technological help from Molecular Vista Inc. Part of experimental work was conducted in Fudan Nanofabrication Lab.

**Author Contributions** Z.A., W.K.L conceived the idea and designed the experiments. W.K.L carried out all experiments with helps from Z.X., L.Q., X.G., H.X.. S.A.B., A.K. and S.C. contributed to theoretical analysis. Y.S. assisted sample fabrication and Z.Z. provided the wafer growth. Z.A. and W.K.L co-wrote the manuscript with comments from all authors. Z.A. and W.L. co-supervised the research project.